\documentclass{elsart}
\usepackage{graphicx}
\usepackage{amsmath}
\begin{document}

\begin{frontmatter}
\author{Luis Din\'{\i}s } and
\author{Juan M.R.~Parrondo}

\address{Grupo Interdisciplinar de Sistemas Complejos (GISC)
and Dept.~de F\'{\i}sica At\'{o}mica,
 Molecular y Nuclear,
 Universidad
Complutense de Madrid, 28040-Madrid, Spain.}

\title{Inefficiency of voting in Parrondo games}

\begin{abstract}
We study a modification of the so-called Parrondo's paradox where
a large number of individuals choose the game they want to play by
voting. We show that it can be better for the players to vote
randomly than to vote according to their own benefit in one turn.
 The former yields a winning tendency while the latter results in steady losses.
  An explanation of
this behaviour is given by noting that selfish voting prevents the
switching between games that is essential for the total capital to
grow. Results for both finite and infinite number of players are
presented. It is shown that the extension of the model to the
history-dependent Parrondo's paradox also displays the same
effect. \end{abstract}

\end{frontmatter}

\section{Introduction}
 The dynamics of a flashing ratchet can be translated into a
counterintuitive phenomenon in gambling games which has recently
attracted considerable attention. It is the so-called {\em
Parrondo's paradox} \cite{stat,nature,fnl} consisting of two
losing games, A and B, that yield, when alternated, a winning
game.

In game A, a player tosses a coin and makes a bet on the throw. He
 wins or loses 1 euro depending on whether the coin falls heads or tails.
 The probability $p_1$ of winning is
$p_1=1/2-\epsilon\text{ with }0\leq\epsilon \ll 1$; so game A is
fair  when $\epsilon=0$ and losing when $\epsilon>0$. By losing,
winning, and fair games here we mean that the average capital is a
decreasing, increasing, and a constant function of the number of
turns, respectively.

\begin{figure}
\begin{center}
\includegraphics[width=10cm]{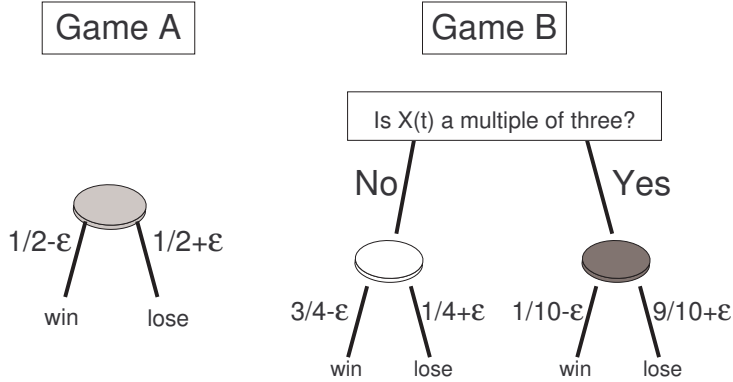}
\caption{Rules of the two Parrondo games}
 \label{fig:rules}
\end{center}
\end{figure}

The second game ---or game B--- consists of two coins. The player
must throw coin 2 if his capital is not a multiple of three, and
coin 3 otherwise. The probability of winning with coin 2 is
$p_2=3/4-\epsilon$ and with coin 3 is $p_3=1/10-\epsilon$. They
are called ``good'' and ``bad'' coins respectively. It can be
shown that game B is also a losing game if $\epsilon>0$ and that
$\epsilon=0$ makes B a fair game \cite{fnl,newpar}. The rules of
both game A and B are depicted in fig. \ref{fig:rules}

Surprisingly, switching between games A and B in a random fashion
or following some periodic sequences produces a winning game, for
$\epsilon>0$ sufficiently small, i.e., the average of player
earnings grows with the number of turns. Therefore, from two
losing games we actually get a winning game. This indicates that
the alternation of stochastic dynamics  can result in a new
dynamics, which differs qualitatively from the original ones.

Alternation is either periodic or random in  the flashing rachet
and in the paradoxical games. On the other hand, we have recently
studied the case of a {\em controlled} alternation of games, where
information about the state of the system can be used to select
the game to be played with the goal of maximising the capital
\cite{dinis}. This problem is trivial for a single player: the
best strategy is to select game A when his capital is a multiple
of three and B otherwise. This yields higher returns than any
periodic or random alternation. Therefore, choosing the game as a
function of the current capital presents a considerable advantage
with respect to ``blind'' strategies, i.e., strategies that do not
make use of any information about the state of the system, as it
is the case of the periodic and random alternation. Also, in a
flashing ratchet, switching on and off the ratchet potential
depending on the location of the Brownian particle allows one to
extract energy from a single thermal bath, in apparent
contradiction with the second law of thermodynamics \cite{review}.
This is nothing but a Maxwell demon, who operates having at his
disposal information about the position of the particle; and it is
the acquisition or the subsequent erasure of this information what
has an unavoidable entropy cost \cite{leff}, preventing any
violation of the second law.

Whereas a controlled alternation of games is trivial for a single
player, interesting and counter-intuitive phenomena can be found
in {\em collective} games. We have recently considered  a
collective version of the original Parrondo's paradox.
 In this model, the game ---A or B--- that a
large number $N$ of individuals play can be selected at every
turn. It turns out that blind strategies are winning whereas a
strategy which chooses the game with the highest average return is
losing \cite{dinis}.

In this paper, we extend our investigation of controlled
collective games considering a new strategy based on a majority
rule, i.e., on voting. This type of rule is relevant in several
situations, such as the modelling of public opinion
\cite{galam,krap} or the design of multi-layer neural networks by
means of {\em committee machines} \cite{barkai,cris}. We will show
that, in controlled games, the rule is very inefficient: if every
player votes for the game that gives him the highest return, then
the total capital decreases, whereas blind strategies generate a
steady gain. The same effect can be found for the
capital-independent games introduced in \cite{newpar}. As
mentioned above, for a single player, the majority rule does
defeats the blind strategies. The inefficiency of voting is
consequently a purely collective effect.

The paper is organised as follows. In Section \ref{sec:model} we
present the model and the counter-intuitive performance of the
different strategies. In Section \ref{sec:analysis} we discuss and
provide an intuitive explanation of this behaviour. In Sec.
\ref{sec:finite}, we analyse how the effect depends on the number
of players. In Section \ref{sec:history}, we extend these ideas to
the capital-independent games introduced in \cite{newpar}.
Finally, in Sec. \ref{sec:conclusions} we present our main
conclusions.

\section{The model}
\label{sec:model}

The model consists of a large number $N$ of players. In every
turn, they have to choose one of the two original Parrondo games,
described in the Introduction and in fig. \ref{fig:rules}. Then
{\em every} individual plays the selected game against the casino.

We will consider three strategies to achieve the collective
decision. {\em a)} The {\em random strategy}, where the game is
chosen randomly with equal probability. {\em b)}  The {\em
periodic strategy}, where the game is chosen following a given
periodic sequence. The sequence that we will use throughout the
paper is $ABBABB\dots$ since it is the one giving the highest
returns. {\em c)} The {\em majority rule (MR) strategy}, where
every player votes for the game giving her the highest probability
of winning, with the game obtaining the most votes being selected.

The model is related to other extensions of the original Parrondo
games played by an ensemble of players, such as those considered
by Toral \cite{toral,toralcapred}. However, in our model the only
interaction among players can occur when the collective decision
is made. Once the game has been selected, each individual plays,
in a completely independent way, against the casino. Moreover, in
the periodic and random strategies there is no interaction at all
among the players, the model being equivalent to the original
Parrondo's paradox with a single player.

\begin{figure}
\begin{center}
\includegraphics[width=7cm]{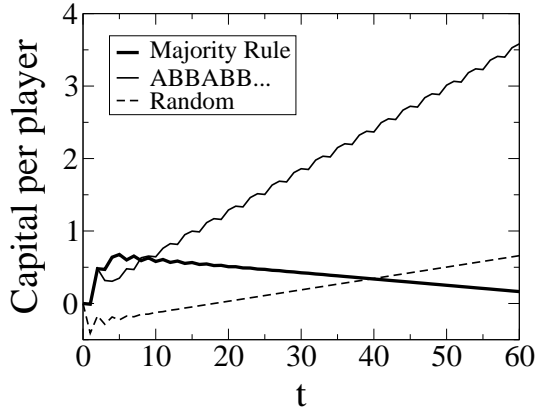}
\caption{Evolution of the capital per player in an infinite
ensemble for $\epsilon=0.005$ and the three strategies discussed
in the text. } \label{capital}
\end{center}
\end{figure}

The MR makes use of the information about the state of the system,
whereas the periodic and random strategies are blind, in the sense
defined above. One should then expect a better performance of the
MR strategy. However, it turns out that, for large $N$, these
blind strategies produce a systematic winning whereas the MR
strategy is losing. This is shown in figure \ref{capital}, where
the capital per player as a function of the number of turns is
depicted for the three strategies and an infinite number of
players (see Appendix \ref{app} for details on how to obtain fig.
\ref{capital}).

\section{Analysis}
\label{sec:analysis}

How many players vote for each game? The key magnitude to answer
this question and  to explain the system's behaviour is
$\pi_0(t)$, the fraction of players whose money is a multiple of
three at turn $t$.  This fraction $\pi_0(t)$ of players vote for
game A in order to avoid the bad coin in game B. On the other
hand, the remaining fraction $1-\pi_0(t)$ vote for game B to play
with the good coin. Therefore, if $\pi_0(t)\geq 1/2$, there are
more votes for game A and, if $\pi_0(t)< 1/2$, then game B is
preferred by the majority of the players.

Let us focus now on the behaviour of $\pi_0(t)$ for $\epsilon=0$
when playing both games separately. If game A is played a large
number of times, $\pi_0(t)$ tends to 1/3 because the capital is a
symmetric and homogenous random walk under the rules of game A. On
the other hand, if B is played repeatedly, $\pi_0(t)$ tends to
5/13. This can be proved by analyzing game B as a Markov chain
\cite{fnl,newpar}. It is also remarkable that, for
$\pi_0(t)=5/13$, the average return when game B is played is zero.

\begin{figure}
\begin{center}
\includegraphics[width=7cm]{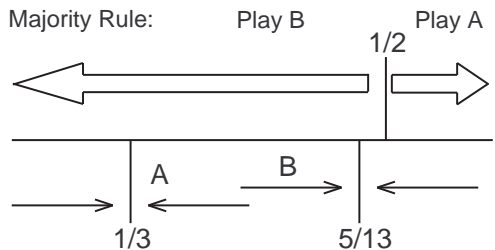}
\caption{Schematic representation of the evolution of $\pi_0(t)$
under the action of game A and game B. The prescription of the MR
is also represented.}\label{scheme}
\end{center}
\end{figure}

 Figure  \ref{scheme} represents schematically the evolution of
$\pi_0(t)$ under the action of each game, as well as the
prescription of the MR strategy explained above.
Now we are ready to
explain why the MR strategy yields worse results than
the periodic and  random sequences.

\begin{figure}
\begin{center}
\includegraphics[width=13cm]{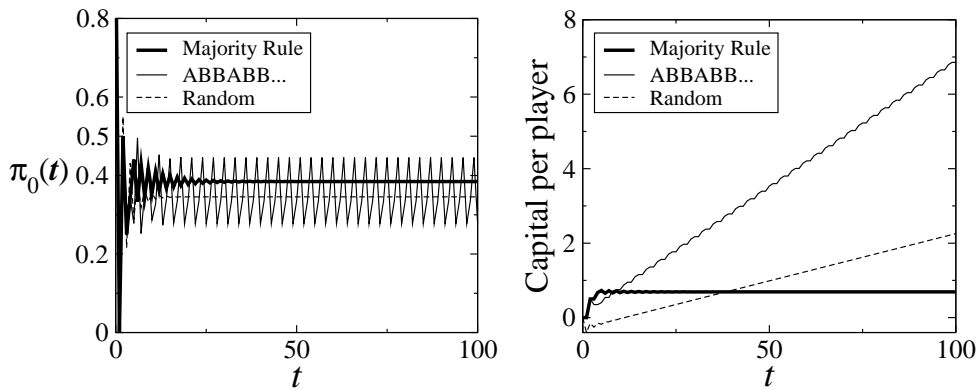}
\caption{Evolution of $\pi_0(t)$ (left) and the capital per player
(right) for $N=\infty$, $\epsilon=0$ for the MR and random
strategies. The MR chooses game B when $\pi_0$ is below the
straight
 line depicted at 1/2 and game A otherwise.
} \label{mayoriae0} \end{center}
\end{figure}

 We see that, as long as
$\pi_0(t)$ does not exceed $1/2$, the MR strategy chooses game B.
However, playing B takes $\pi_0$ closer to $5/13$, well below
$1/2$, and thus more than half of the players vote for game B
again.
 After a number
of runs, the MR strategy gets trapped playing game B forever. Then
$\pi_0$ asymptotically approaches 5/13, and as this happens, game
B turns into a fair game when $\epsilon=0$. As a consequence, the
MR will not produce  earnings any more, as can be seen in figure
\ref{mayoriae0}.

The introduction of $\epsilon>0$ turns game B into a losing game
if played repeatedly. Consequently, the MR strategy becomes a
losing one as in figure \ref{capital}.   To overcome this losing
tendency, the players must sacrifice their short-range profits,
not only for the benefit of the whole community but also for their
own returns in the future. Hence, some kind of cooperation among
the players is needed
 to prevent them from losing their capital. A similar effect has been found by Toral in
 another version of collective Parrondo's games. There, sharing the capital among
 players induces a steady gain \cite{toral}.
In our case, the striking result is that no complex cooperation is
necessary. It is enough that the players agree to vote at random.

\section{Finite number of players}
\label{sec:finite}

 In the previous analysis an infinite number of players has been
considered. Remarkably, for just one player the
 MR strategy trivially performs
better than any periodic or random sequence, since it completely
avoids the use of the bad coin. In this Section we analyse the
crossover between the winning behaviour for a small number of
players and the losing behaviour when this number is large.

\begin{figure}
\begin{center}
\includegraphics[width=10cm]{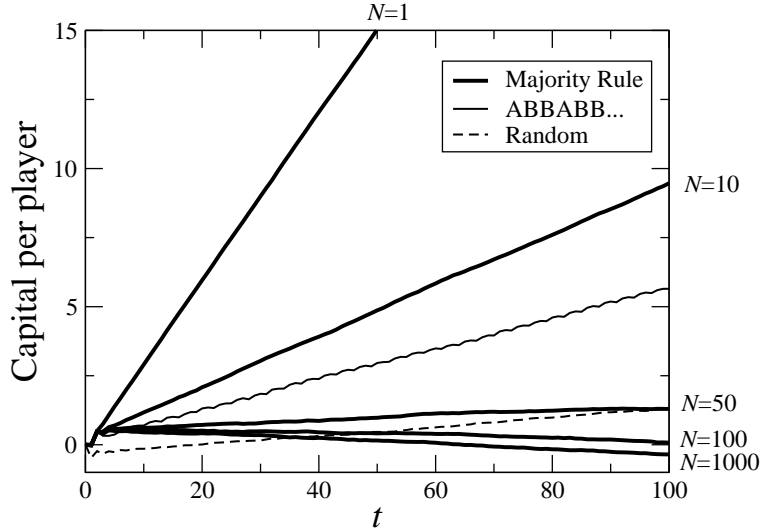}
\caption{Simulation results for the average capital per player for
$N=10$, 50, 100, and 1000 players, $\epsilon=0.005$, and the three
different  strategies. The simulations have been made over a
variable number of realizations, ranging from 10000 realizations
for $N=1$ to 10 realizations for $N=1000$. Simulations for the
random and periodic strategies have been made with $N=100$ players
and averaging over 100 realizations. For these blind strategies,
the result does not depend on the number of players $N$.}
\label{figura5}
\end{center}
\end{figure}

Figure \ref{figura5} shows numerical results of the average
capital per player for an increasing number of players ranging
from 10 to 1000.
One can observe that, the larger the number of players, the worse
the results for the MR strategy, becoming losing for a number of
players between 50 and 100.

The above discussion for an infinite ensemble allows us to give a
qualitative explanation. The difference between large and small
$N$ is the magnitude of the fluctuations of $\pi_0(t)$ around its
expected value. If game B is chosen a large number of times in a
row, then the expected value of $\pi_0(t)$ is 5/13. On the other
hand, the MR selects B unless $\pi_0(t)$ is above 1/2. Therefore,
for the MR to select A, fluctuations must be of order
$1/2-5/13=3/26\simeq 0.115$.  For $N$ players, the fraction of
players with capital multiple of three, $\pi_0(t)$, will be a
random variable following a binomial distribution, at least if B
has been played a large number of times in a row. If the expected
value of $\pi_0(t)$ is 5/13, fluctuations of $\pi_0(t)$ around
this value are of order $\sqrt{5/13\times 8/13\times 1/N}$. Then,
fluctuations will allow the MR strategy to choose A if $N\simeq
20$. Far above this value, fluctuations that drive $\pi_0(t)$
above 1/2 are very rare, and MR chooses B at every turn. On the
other hand, for $N$ around or below 20, there is an alternation of
the games that can even beat the optimal periodic strategy.

We see that the MR strategy can take profit of fluctuations much
better than blind strategies, but it loses all its efficiency when
these fluctuations are small. We believe that this is closely
related to the second law of thermodynamics. The law prohibits any
decrease of entropy only in the thermodynamic limit or for average
values. On the other hand, when fluctuations are present, entropy
can indeed decrease momentarily and this decrease can be exploited
by a Maxwell demon.

\section{History dependent games}
\label{sec:history}

A similar phenomenon is exhibited by the games introduced in Ref.
\cite{newpar}, whose rules depend on the history rather than on
the capital of each player. Game A is still the same as above,
whereas game B is played with three coins according to the
following table:
\begin{center}
\begin{tabular}{c|c|c|c}
 Before last & Last&  Prob. of win & Prob. of loss \\
$t-2$        & $t-1$ &   at $t$ & at $t$ \\\hline loss & loss &
$p_1$ & $1-p_1$\\ loss & win   & $p_2$ & $1-p_2$
\\ win & loss  & $p_2$ & $1-p_2$ \\ win & win &
 $p_3$ & $1-p_3$ \\
\end{tabular}
\end{center}
with $p_1=9/10-\epsilon$, $p_2=1/4-\epsilon$, and
$p_3=7/10-\epsilon$.

\begin{figure}
\begin{center}
\includegraphics[width=8cm]{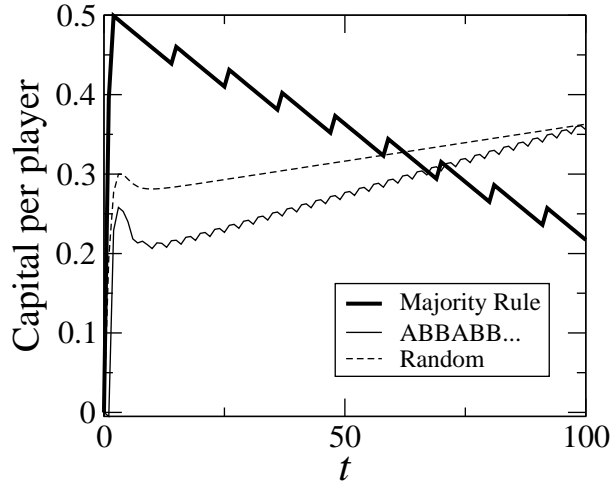}
\caption{Evolution of the average money of the players in the
history-dependent games for  $\gamma=0.5$, $\epsilon=0.005$ and
 three different
strategies.} \label{histo} \end{center}
\end{figure}

Introducing a large number of players but allowing just a randomly
selected fraction $\gamma$ of them to vote and play, the same ``voting paradox''
is recovered for sufficiently small $\gamma$. Again, blind strategies achieve a
constant growth of the average capital with the number of turns while the  MR
strategy returns a decreasing average capital, as it is shown in figure
\ref{histo}.

\section{Conclusions}
\label{sec:conclusions}

We have shown that the paradoxical games based on the flashing
ratchet exhibit a counterintuitive phenomenon when a large number
of players are considered. A majority rule based on selfish voting
turns to be very inefficient for large ensembles of players. We
have also discussed how the rule only works for a small number of
players, since in that case it is able to exploit capital
fluctuations.

The interest of the model presented here is threefold. First of
all, it shows that cooperation among individuals can be beneficial
for everybody. In this sense, the model is related to that
presented by Toral in Ref. \cite{toralcapred}. Since John Maynard
Smith first applied game theory to biological problems
\cite{maynard}, games have been used in ecology and social
sciences as models to explain social behaviour of individuals
inside a group. Some generalizations of the voting model might be
useful for this purpose. For instance, it could be interesting to
analyse the effect of mixing selfish and cooperative players  or
the introduction of players who could change their behaviour
depending on the fraction of selfish voters in previous turns.

Secondly, the effect can also be relevant in random decision
theory or the theory of stochastic control \cite{white} since it
shows how periodic or random strategies can be better than some
kind of optimization. In this sense, there has been some work on
general adaptive strategies in games related with Parrondo's
paradox \cite{behrends,rahmann}.

Thirdly, this model and, in particular, the analysis for $N$
finite, prompts the problem of how information can be used to
improve the performance of a system. In the models presented here,
information about the fluctuations of the capital is useful only
for a small number of players, that is, when these fluctuations
are significant. It will be interesting to analyse this crossover
in further detail, not only in the case of the games but also for
Brownian ratchets. Work in this direction is in progress.

\begin{ack}
The authors gratefully acknowledge fruitful discussions with
Christian Van den Broeck, who suggested us the introduction of the
MR strategy. We also thank H. Leonardo Mart\'{\i}nez for valuable
comments on the manuscript.
 This work has been financially supported by grant
BFM2001-0291-C02-02 from Ministerio de Ciencia y Tecnolog\'{\i}a
(Spain) and by a grant from the {\em New del Amo Program}
(Universidad Complutense).
\end{ack}

\appendix
\section{Evolution equations}
\label{app}

In this Section we describe the semi-analytical solution of the
model for an infinite number of players, used to depict fig.
\ref{capital}. Let $\pi_i(t)$, be the fraction of players whose
capital at turn $t$ is of the form $3n+i$ with $i=0,1,2$ and $n$
an integer number.

If game A is played in turn $t$, these fractions change following
the expression \cite{newpar}:
\begin{equation} \left(
\begin{array}{c} \pi_0(t+1) \\ \pi_1(t+1) \\ \pi_2(t+1)\end{array}\right)=\left(
\begin{array}{ccc} 0 &\quad 1/2+\epsilon\quad  & \quad 1/2-\epsilon\quad \\
\quad 1/2-\epsilon\quad & 0 & 1/2+\epsilon \\
1/2+\epsilon & 1/2-\epsilon & 0
\end{array} \right) \left(
\begin{array}{c} \pi_0(t) \\ \pi_1(t) \\
\pi_2(t)\end{array}\right) \label{matrizA}
\end{equation}
which can be written in a vector notation as:
\begin{equation}
\overrightarrow{\pi}(t+1)=\Pi_A\overrightarrow{\pi}(t).
\end{equation}
Similarly, when B is played, the evolution is given by:
\begin{equation}
\overrightarrow{\pi}(t+1)=\Pi_B\overrightarrow{\pi}(t)
\end{equation}
with
\begin{equation} \Pi_B=\left(
\begin{array}{ccc} 0 &\quad 1/4+\epsilon\quad &\quad 3/4-\epsilon\quad\\
\quad 1/10-\epsilon\quad & 0 & 1/4+\epsilon \\ 9/10+\epsilon &
3/4-\epsilon & 0
\end{array} \right).
\end{equation}

Now we can write the evolution equation for each strategy. For the
{\em random strategy}: \begin{equation}
\overrightarrow{\pi}(t+1)=\frac{1}{2}\left[\Pi_A+\Pi_B\right]\overrightarrow{\pi}(t).
\end{equation}

For the {\em periodic strategy} (ABBABB..):
\begin{equation}
\overrightarrow{\pi}(3(t+1))=\Pi_B^2\Pi_A\overrightarrow{\pi}(3t).
 \end{equation}

Finally, with the MR strategy the ensemble plays game A if
$\pi_0(t)\ge 1/2$ and   B otherwise. Therefore:
\begin{equation}
 \overrightarrow{\pi}(t+1)=\left\{
\begin{array}{ll}
\Pi_A\overrightarrow{\pi}(t) & \mbox{if $\pi_0(t)\ge 1/2$} \\
\Pi_B\overrightarrow{\pi}(t) & \mbox{if $\pi_0(t)< 1/2$.}
\end{array}\right.
\end{equation}

Notice that  the MR strategy is the only one inducing a nonlinear
evolution in the population fractions. To calculate the evolution
of the capital, we compute the winning probability in each game:
\begin{eqnarray}
p_{\rm win}^A(t)&=& \frac{1}{2}-\epsilon \nonumber \\
p_{\rm win}^B(t)&=& \frac{1}{10}\pi_0(t)+\frac{3}{4}\left(
1-\pi_0(t)\right)-\epsilon.
\end{eqnarray}
 Finally, the average capital
  $\langle X(t)\rangle $ per player evolves as:
\begin{equation} \langle X(t+1)\rangle  =\langle  X(t)\rangle  + 2p_{\rm
win}(t)-1
\end{equation}
and $p_{\rm win}(t)$ is replaced by $p_{\rm win}^A(t)$ or $p_{\rm
win}^B(t)$, depending on the game played at turn $t$ in each
strategy.

 %--------------------REFERENCIAS-----------------------------

\end{document}